\documentclass[aps,prx,showpacs,  amsmath,amssymb,longbibliography,showkeys,12pt]{revtex4-2}%
\usepackage{amsfonts}
\usepackage{amsmath}
\usepackage{amssymb}
\usepackage{graphicx}
\usepackage{epsfig}
\usepackage{exscale}
\usepackage{float}
\usepackage{bm}
\usepackage{suffix}
\usepackage{mathtools}
\usepackage{newunicodechar}
\usepackage{lipsum}
\usepackage{enumitem}
\usepackage{bbding}
\usepackage{tikz}
\usepackage{multirow}%
\usepackage[colorlinks=true,linkcolor=blue]{hyperref}
\usepackage{ifpdf }
\usepackage{silence}
\expandafter\ifx\csname package@font\endcsname\relax\else
 \expandafter\expandafter
 \expandafter\usepackage
 \expandafter\expandafter
 \expandafter{\csname package@font\endcsname}
\fi
\providecommand{\U}[1]{\protect\rule{.1in}{.1in}}

\DeclarePairedDelimiterX\MeijerM[3]{\lparen}{\rparen}
{\begin{smallmatrix}#1 \\ #2\end{smallmatrix}\delimsize\vert\,#3}
\newcommand\MeijerG[8][]{  G^{\,#2,#3}_{#4,#5}\MeijerM[#1]{#6}{#7}{#8}}
\WithSuffix
\newcommand\MeijerG*
[7]{
G^{\,#1,#2}_{#3,#4}\MeijerM*{#5}{#6}{#7}}

\begin{document}
\title[ ]{ Toward Exact Critical Exponents from the low-order loop expansion of the Effective
Potential in Quantum Field Theory }
\author{ Abouzeid M. Shalaby}
\email{amshalab@qu.edu.qa}
\affiliation{ Physics and Materials Sciences  Department, College of Arts and Sciences, Qatar University, P.O box 2713, Doha, Qatar.}
\begin{abstract}
 The asymptotic strong-coupling behavior as well as the exact critical exponents from   scalar field theory even for the simplest case of $1+1$  dimensions have not been obtained yet.  Hagen Kleinert  has linked both critical exponents and strong coupling parameters to each other.  He used a  variational technique ( back to kleinert and Feynman)  to extract accurate values for the strong coupling parameters from which he was able to extract  precise critical exponents. In this work,  we suggest a simple method of using  the effective potential ( low order)  to obtain exact values for the strong-coupling parameters for the $\phi^{4}$ scalar field theory in $0+1$ and $1+1$ space-time dimensions. For the  $0+1$ case, our results coincide with the well-known exact values already known from literature while  for the $1+1$ case we test the results by obtaining the corresponding exact critical exponent. As the effective potential  is a well-established tool in quantum field theory, we expect that the results can be easily extended to the most important three dimensional case and then the dream of getting exact critical exponents is made possible.   
\end{abstract}
\maketitle
Critical phenomena does exist in almost every branch in physics. The so far
inviolate conjecture of the existence of universality classes
\cite{Hollowood2013} makes the subject to be studied from different point of
views as interaction details do not matter near the critical region for a
system undergoing a second-order phase transition. For instance, one might
find a model from fluids that lies in the same class of universality with a
model from magnetism or a field theoretic model. As an example, the critical
exponents associated with $^{4}He$ superfluid phase transition coincides with
that of the $O\left(  2\right)  -$ symmetric $\phi^{4}$ scalar field model
\cite{Kleinert-Borel,abodispute,Abo-expon7}. The phenomena is completely
non-perturbative and thus one always needs to resort to non-perturbative
techniques to extract reliable results.

Different computational tools have been used in literature to study second
order phase transition. Out of these, there is the Bootstrap conformal field
theory (CFT) \cite{dispute,dispute1}, high temperature (HT) expansion
\cite{Butera2002, Butera-PRB}, Monte Carlo simulation \cite{MC20N3,MC19L,MC19}
and renormalization group in quantum field theory \cite{Kleinert1995,
Kleinert-Borel, Adzhemyan2019, Abo-expon7, zin-cr,
Borel-6L,Schnetz2018,LeGuillou1977,abo-precize}. In two dimensions, however,
the Ising model has been solved exactly \cite{Onsager} and thus exact critical
exponents are already known \cite{Nienhuis}. In three dimensions, on the other
hand, exact values for critical exponents have not been obtained yet by any means.

In literature, there exists what is known by the $\lambda$-point dispute where
the theoretical and experimental predictions for the specific heat exponent
$\alpha$ of the $^{4}He$ superfluid phase transitions do not \ match
\cite{dispute,dispute1,dispute1,MC19,experiment}. Accordingly, the need for
exact results is very important in order to decide if the experiment has to be
repeated or to be confirmed with such aimed exact results.

Like the other computational tools for the calculation of critical exponents
in three dimensions, the renormalization group method in quantum field theory
( so far) is not able to predict exact results. Even more and up to the best
of our knowledge, such technique is not able to predict exact critical
exponents even for the two dimensional case. In fact, in three dimensions, the
best results from renormalization group have been obtained from our previous
calculations in Refs. \cite{Abo-expon7,abodispute} where we treated the
seven-loop order using hypergeometric-Meijer resummation algorithm.

In quantum field theory, the strong-coupling asymptotic behavior is not known
as well. It seems that both the unknown exact critical exponents in quantum
field theory and the unknown parameters that defines the strong-coupling
behavior are related. In fact, Hagen Kleinert has shown that near the critical
point , a renormalization group function $Q\left(  g_{0}\right)  $ of the bare
coupling $g_{0}$ can be expanded as
\cite{Kleinert-strong1,Kleinert-Borel,keinert-strong,variational95,Janke95}
\begin{equation}
Q\left(  g_{0}\right)  =Q\left(  g^{\ast}\right)  +\acute{Q}\left(  g^{\ast
}\right)  \times\frac{const}{g_{0}^{\frac{\omega}{\varepsilon}}}+.....,
\label{strong-q}%
\end{equation}
where \ $\omega$ is the approach to scaling critical exponent and
$\varepsilon=4-d$, where $d$ is the dimension of the space-time. In general,
the strong coupling expansion takes the form
\begin{equation}
f\left(  g_{0}\right)  =g_{0}^{\frac{p}{q}}\sum_{i=0}^{\infty}c_{i}\left(
g_{0}^{\frac{-2}{q}}\right)  ^{i}, \label{strong-qq}%
\end{equation}
however, the IR fixed point forced $p$ to be zero for all renormalization
group functions \cite{Kleinert-strong1,Kleinert-Borel,keinert-strong}. The
most important point here is that $p$ and $q$ are playing the main rule in the
prediction of the renormalization group functions as $g_{0}\rightarrow\infty$
( critical point). In Refs.\cite{Janke95,
Kleinert-strong1,Kleinert-Borel,keinert-strong}, the authors used a kind of
variational technique where the perturbative coefficients (weak-coupling) are
used as input to predict accurate values for $p$ and $q$ from which one can
extract critical exponents. Using this technique, they were able to get
accurate (approximate) results for the critical exponents in $d=4-\epsilon$ dimensions.

As we mentioned above, the exact critical exponents as well as the exact $p$
and $q$ parameters are not known so far even for a simple field theory like
the $\phi^{4}$ model and even for the low two dimensional case. Seeking an
algorithm to find the parameters $p$ and $q$ will lead certainly to the first
exact critical exponent in three dimensions in general and the first exact
critical exponent from field theory ( not CFT) in two dimensions. In this
work, we \ suggest a simple algorithm to predict the values of $p$ and $q$ for
a quantum field theory. The applications we list in this work include two
examples: first, we consider \ the case of $0+1$ dimensions (quantum
mechanics) where exact values are known and thus can be compared with our
results. Second, we consider the $1+1$ dimensions and obtain exact values for
$p$ and $q$ which are totally new results and can be tested only by obtaining
the corresponding critical exponents and compare them with the already known
exact ones from the Ising model. In fact and up to the best of our knowledge,
it is also the first time to obtain exact critical exponents from QFT in two
dimensions. Note that $\frac{2}{q}=\frac{\omega}{\varepsilon}\ $and thus
knowing $q$ will certainly determine $\omega$.

\begin{figure}
\begin{center}
\includegraphics[width=10cm,height=7cm]{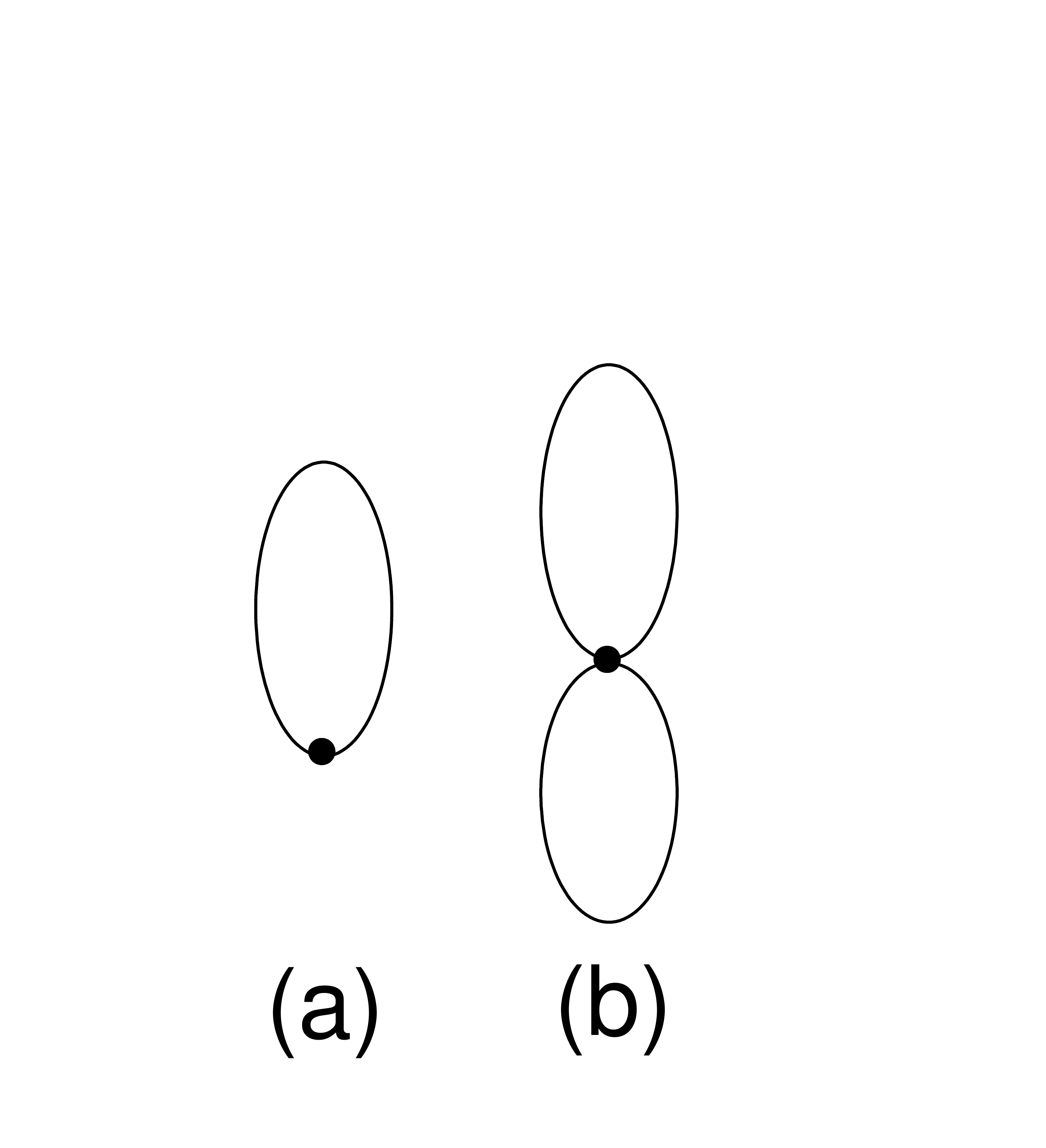}
\end{center}
\caption{\textit{The $1$-vertex Feynman diagrams contributing to the  vacuum energy for the $\phi^{4}$ field theory.}}
\label{Feyn-cact2}
\end{figure}

To start, consider the Hamiltonian density of the one component $\phi^{4}$
scalar field theory:%
\begin{equation}
H=\left(  \frac{1}{2}\nabla\phi^{2}+\frac{1}{2}\pi^{2}+\frac{1}{2}m^{2}%
\phi^{2}+\frac{\lambda}{4}\phi^{4}\right)  ,
\end{equation}
where $\pi=\dot{\phi}$. In $0+1$ space-time dimensions, the Hamiltonian
takes the form:%
\begin{equation}
H=\frac{1}{2}\pi^{2}+\frac{1}{2}m^{2}\phi^{2}+\frac{\lambda}{4}\phi^{4}.
\end{equation}
The effective field can be introduced via the canonical relations%
\[
\phi=\psi+v,\text{ \ }\pi=\Pi=\dot{\psi},\text{\ \ \ }%
\]
where $v$ is the vacuum expectation value ($vev$). This transformation leads the Hamiltonian operator to take the form:%
\[
H=H_{0}+H_{I}+\frac{1}{2} m^2 v^2+\frac{\lambda}{4}v^{4},
\]
where for the simplest case of $m=0$ ( equivalent to strong coupling)$,$ we
have:
\begin{align}
H_{0} &  =\frac{1}{2}\left(  \left(  \nabla\psi\right)  ^{2}+\Pi^{2}+M^{2}%
\psi^{2}\right)  ,\nonumber\\
\ H_{I} &  =\frac{\lambda}{4}\left(  \psi^{4}+4v\psi^{3}\right)  +\left(  -\frac
{1}{2}M^{2}+\frac{3}{2}\lambda v^{2}\right)  \psi^{2}+\lambda v^{3}\psi.
\end{align}
$M$ here is the mass of the field $\psi$. The vacuum energy or equivalently
the effective potential\ $V_{eff}\left(  v\right)  $ is given by;%
\begin{equation}
V_{eff}\left(  v\right)  =\langle0|H|0\rangle=\langle0|H_{0}+H_{I}+\frac{\lambda}%
{4}v^{4}|0\rangle.
\end{equation}
The effective potential is the generating functional for the 1- particle
irreducible amplitudes \cite{Peskin} and thus we have the relations;
\begin{equation}
\frac{\partial V_{eff}}{\partial B}=0\text{, }\frac{\partial^{2}V_{eff}%
}{\partial B^{2}}=M^{2}.\label{paramf}%
\end{equation}
Taking only the lowest order (1-vertex) Feynman diagrams in Fig.\ref{Feyn-cact2} into account, we get the
results:%
\begin{equation}
E_{0}=\frac{1}{4}M+\frac{3}{16}\frac{\lambda}{M^{2}}+\frac{3}{4M}\lambda v^{2}+\frac{\lambda}%
{4}v^{4}.
\end{equation}
Note that this is exactly the Gaussian effective potential result ( Eq.(2.12) in
Ref.\cite{GEPQM} for $m=0$). As it was  stated in Ref.\cite{GEPQM}, this result includes the one-loop effective potential \cite{Peskin}. In fact, if we drop the two-loop contribution which is the second term in the above equation ( contribution from graph $b$ in Fig. \ref{Feyn-cact2}) we get the one-loop effective potential listed in literature. 

The stability conditions in Eq.(\ref{paramf}) result in:%
\begin{align}
\lambda v^{3}+\frac{3\lambda}{2M}v &  =0,\nonumber\\
\lambda v^{2}+\frac{3}{2M}\lambda  &  =M^{2}.
\end{align}
It has been  stated in Ref.\cite{GEPQM} that the minimum of the effective potential exists at $v=0$. In fact, our equations lead  also to the solution;
\[
v=0,\text{ }M=\frac{1}{2}\sqrt[3]{12}\sqrt[3]{ \lambda}\text{ and }V_{eff}=\frac
{3}{16}\sqrt[3]{12}\sqrt[3]{ \lambda}%
\]
 Note that the $\lambda ^{\frac{1}{3}}$ asymptotic behavior we obtained
here coincides with the exact behavior $V_{eff}\left(  v\right)  \propto
 \lambda^{\frac{1}{3}}$ known for the $x^{4}$ anharmonic oscillator
\cite{harm-sigma,Large-strong1}. So it seems that the simple low-order
effective potential knows about the strong coupling behavior. 

There is another broken-symmetry solution ($v  \neq 0$) for this theory. However, in this case the theory is non-Hermitian ( $\mathcal{PT}$-symmetric) for which we have shown in Ref.\cite{x4pabo} that it has a stable effective potential for $v\neq 0$. In taking   into account the equivalent Hermitian form obtained in Ref.\cite{Jonesx4}, which is equivalent to the original Hermitian one for strong couplings, we can reach the same asymptotic behavior via the symmetry breaking solution. 

We considered above the $m=0$ case  that gives the exact ratio $\frac{p}{q}=\frac{1}{3}$, which suggests
that in expanding the effective potential around $m=0$, one might aim to get
the exact values for both $p$ and $q$. For $m\ne0$,   we have
\begin{equation}
V_{eff}\left(  v\right)  =\frac{1}{4}M+\frac{3}{16}\frac{ \lambda}{M^{2}}+\left(
\frac{1}{2}m^{2}+\frac{3}{2} \lambda v^{2}\right)  \frac{1}{2M}+\frac{1}{2}m^{2}%
v^{2}+\frac{ \lambda}{4}v^{4},\nonumber
\end{equation}
where the stability conditions give
\begin{align*}
\frac{3}{2}g\frac{v}{M}+m^{2}v+ \lambda v^{3}  & =0\\
\frac{3}{2}\frac{ \lambda}{M}+m^{2}+3 \lambda v^{2}  & =M^{2}%
\end{align*}
For $v=0,$ $\frac{3}{2}\frac{ \lambda}{M}+m^{2}=M^{2}$ we have
\begin{align*}
M  & =\frac{1}{6}\sqrt[3]{162 \lambda+6\sqrt{-48m^{6}+729 \lambda^{2}}}+2\frac{m^{2}%
}{\sqrt[3]{162 \lambda+6\sqrt{-48m^{6}+729 \lambda^{2}}}}\\
V_{eff}\left(  0\right)    & =\frac{1}{4}M+\frac{3}{16}\frac{ \lambda}{M^{2}}%
+\frac{1}{4}\frac{m^{2}}{M}%
\end{align*}
The expansion of $V_{eff}$ around $m^{2}$ shall give the following
series :%
\[
V_{eff}\left(  v\right)  = \lambda^{\frac{1}{3}}\left(  c_{0}+c_{1} \lambda^{\frac{-2}{3}%
}+c_{3} \lambda^{\frac{-4}{3}}+....\right)  .
\]
This result gives $\frac{p}{q}=\frac{1}{3}$ and $\frac{2}{q}=\frac{2}{3}$ or
$p=1$ and $q=3$. These values for $p$ and $q$ coincide with the known exact
values \cite{Large-strong1}. Also, when we apply the same technique to the
$\phi^{6}\ $theory in $0+1$ dimensions we get the result
\[
V_{eff}\left(  v\right)  = \lambda^{\frac{1}{4}}\left(  c_{0}+c_{1} \lambda^{\frac{-2}{4}%
}+c_{3} \lambda^{\frac{-4}{4}}+....\right)  .
\]
with $p=1$ and $q=4$ which again are exact \cite{Large-strong1}

For the $\phi^{4}$ theory in $1+1$ space-time dimensions, the  effective
potential is divergent. So one needs to employ the well known renormalization
process followed by scale invariance of bare quantities. In fact, the
divergences associated with the Feynman diagrams in Fig. \ref{Feyn-cact2}  can be absorbed
using only normal ordering \cite{normalc,Shalaby2007}. So let us ( without
loss of generality) start with a normal-ordered Hamiltonian density of the
form;%
\begin{equation}
H=N_{m}\left(  \frac{1}{2}\nabla\phi^{2}+\frac{1}{2}\pi^{2}+\frac{1}{2}%
m^{2}\phi^{2}+\frac{\lambda}{4}\phi^{4}\right),%
\end{equation}
where $N_{m}$ stands for normal ordered fields with respect to the mass
parameter $m$ and\ $\pi=\frac{\partial\phi}{\partial t}$. One can rewrite the
normal ordered different terms ($N_{m})$ in terms of same quantities but
normal ordered with respect to another mass parameter $M=\sqrt{t}m$ in using
the relation \cite{normalc,normal2}
\begin{equation}
N_{m}\exp\left(  i\beta\phi\right)  =\exp\left(  -\frac{1}{2}\beta^{2}%
\Delta\right)  N_{M=\sqrt{t}\cdot m}\exp\left(  i\beta\phi\right)
\text{,}\label{normal2}%
\end{equation}
where
\begin{equation}
\Delta=-\langle0\left\vert \phi(x)\phi(x)\right\vert 0\rangle.
\end{equation}
In $1+1$ dimensions, one can get the normal ordering of the kinetic term as:
\begin{equation}
N_{m}\left(  \frac{1}{2}\left(  \nabla\phi\right)  ^{2}+\frac{1}{2}\pi
^{2}\right)  =N_{M}\left(  \frac{1}{2}\left(  \nabla\phi\right)  ^{2}+\frac
{1}{2}\pi^{2}\right)  +\frac{1}{8\pi}\left(  M^{2}-m^{2}\right)
.\label{norkin1}%
\end{equation}
After a canonical transformation of the form $\phi(x)=\psi(x)+v$, where $v$ is
the vacuum condensate, one can obtain the vacuum energy in the form:%
\begin{equation}
V_{eff}\left(  v\right)  =\frac{m^{2}}{8\pi}\left(  B^{2}-G\left(  \frac{1}%
{4}B^{4}+\frac{3}{4}\ln^{2}t-\frac{3}{2}B^{2}\ln t\right)  +\left(  t-\ln
t-1\right)  \right)  ,\label{Veff}%
\end{equation}
where $B^{2}=4\pi v^{2}$ and the dimensionless parameters $t=\frac{M^{2}%
}{m^{2}}$, $G=\frac{\lambda}{2\pi m^{2}}$. Again this result is in agreement with the Gaussian effective potential result in \cite{GEPII}. The only difference is    divergent  constant $D$ in Eq.(4.8) there where the authors stated that it is of no physical consequences.

In using the stability condition
$\frac{\partial V_{eff}\left(  v\right)  }{\partial v}=0$ and the mass
renormalization condition \ $\frac{\partial^{2}V_{eff}\left(  v\right)
}{\partial v^{2}}=M^{2},$ we get the results (for $B\ne0$ as the $B=0$ solution is trivial):%
\begin{align}
1\ +\frac{1}{2}GB^{2}-\frac{3}{2}G\ln t  &  =0\nonumber\\
1\ +\frac{3}{2}GB^{2}-\frac{3}{2}G\ln t  &  =t
\end{align}
which can be solved to give
\begin{align}
t  &  =\ -3G\operatorname{ W}\left( \frac{-1}{3G}\exp\left(  \frac{2}%
{3G}\right)  \right)  ,\nonumber\\
B^{2}  &  =-3\operatorname{ W}\left( \frac{-1}{3G}\exp\left(  \frac{2}%
{3G}\right) \right) ,
\end{align}
where the Lambert's $W$ function is defined as $W(x)e^{ W(x)}=x$.  Using these
results, one can expand the effective potential in Eq.(\ref{Veff}) around
$m=0$, to get:$\allowbreak$%
\[
V_{eff}\left(  \lambda,m\right)  =\frac{1}{8\lambda}m^{4}+\frac{1}{4}\frac
{\pi}{\lambda^{2}}m^{6}+\frac{1}{3}\frac{\pi^{2}}{\lambda^{3}}m^{8}+O\left(
m^{10}\right)  \ .
\]
This equation takes exactly the form in (\ref{strong-qq}))
\begin{align*}
\frac{V_{eff}\left(  \lambda_{m}\right)  }{m^{2}}  &  =\frac{1}{8\lambda_{m}%
}+\frac{1}{4}\frac{\pi}{\lambda_{m}^{2}}\ +\frac{1}{3}\frac{\pi^{2}}%
{\lambda_{m}^{3}}\ +O\left(  \lambda_{m}^{-4}\right)  \ \\
&  =\lambda_{m}^{\frac{p}{q}}\sum_{i=0}^{\infty}c_{i}\left(  \lambda
_{m}^{\frac{-2}{q}}\right)  ^{i},
\end{align*}
where $\lambda_{m}$ is the dimensionless bare coupling $\frac{\lambda}{m^{2}}%
$. Here $\frac{p}{q}=-1$ and $\frac{-2}{q}=-1$. So our predictions for $p$ and
$q$ are $p=-2$ and $q=2$. For the above examples of anharmonic oscillators our
predictions for $p$ and $q$ coincide with their exact values and the question
now is that, are the $p$ and $q$ values predicted from the effective potential
for the $1+1$ scalar $\phi^{4}$ theory exact? If so they should give the exact
critical exponent $\omega$ which is known exactly to be $2$ from Ising model
and also from conformal field theory. The kleinert form in Eq.(\ref{strong-q})
suggests that $\frac{\omega}{\varepsilon}=\frac{2}{q}$. In $1+1$ dimensions
$\varepsilon=2$ and thus $\omega=\frac{4}{q}=2$, which coincides with the exact
result from Ising model and conformal field theory. This means that our
prediction for the strong-coupling parameters for the theory under
consideration is exact and this result is totally new.

The effective potential technique is very familiar in quantum field theory and
thus the extension to higher dimensions is direct. However, there are more
divergent non-cactus Feynman diagrams for which normal ordering can not
regularize and one needs to employ the recipe for regularization and
renormalization process. Note that after getting the effective potential, it
should be in terms of the dimensionless bare coupling and then expand around
$m=0$, where $m$ can be taken as the renormalized mass \cite{Kleinert-Borel}.
This is our aim in a future work which will take a substantial amount of time.

To conclude, so far the exact asymptotic strong-coupling behavior of the
$\phi^{4}$ scalar field theory has not been obtained before this work even for
the simpler two dimensional case. On the other hand, also exact critical
exponents for the three dimensional case has not been obtained by any
computational technique so far and although Ising critical exponents are
exactly known in two dimensions, the quantum field calculations have not
produced exact results.  In fact, the exact solution for the $\phi^{4}$ theory
has not been obtained yet for any space-time dimensions. Accordingly, the
search for exact critical exponents might go through the asymptotic behaviors
of the theory. In a previous article \cite{universal}, we showed that the
critical exponents can be solely extracted from the large-order parameter of
the strong-coupling expansion for the associated physical quantity. However,
till now this parameter is not known exactly for the respective theory. On the
other hand, Hagen Kleinert showed that critical exponents can be extracted
from the asymptotic strong-coupling behavior of the given theory. However, he
used a variational technique to obtain accurate values for such parameters
from the weak-coupling perturbation series as input.

In quantum field theory, there exists a well established other variational
technique called the effective potential which can probe the broken symmetry
phase as well as its loop expansion can be obtained in a systematic way. Moreover, renormalization
techniques can be applied easily within such computational tool. When we
applied such technique to the massless $x^{n}$ anhrmonic oscillator, we
realized that the simple one-vertex approximation can give exactly the
strong-coupling parameter $s$ where a quantity behaves as $A\  \lambda^{s}$ when
$ \lambda\rightarrow\infty$. So we suggested to expand the one-loop result around
$m=0$ and to watch the behavior of the expansion. We realized that it is of
the form in Eq.(\ref{strong-qq}) with exact values for the parameters $p$ and
$q$. Since it is direct to extend it to higher dimensions, we applied the idea
to the $\phi_{1+1}^{4}$ scalar field theory and tested the results by
obtaining the approach to scaling critical exponent $\omega$. We obtained the
result $\omega=2$, in compelete agreement with exact results known from the
square-lattice Ising model.

The effective action is the generating functional of the $1P$-irreducible
amplitudes \cite{Peskin} and thus one can obtain the respective amplitudes by
successive differentiation of the effective potential. Besides, its
calculation in higher dimensions are known which needs field, mass and
coupling renormalization. So we expect that the low order renormalized
effective potential in three dimensions can produce exact strong-coupling
parameters from which one will be able to extract the first exact critical exponent.

\bibliography{Exact-Strong}

\begin{thebibliography}{36}%
\makeatletter
\providecommand \@ifxundefined [1]{%
 \@ifx{#1\undefined}
}%
\providecommand \@ifnum [1]{%
 \ifnum #1\expandafter \@firstoftwo
 \else \expandafter \@secondoftwo
 \fi
}%
\providecommand \@ifx [1]{%
 \ifx #1\expandafter \@firstoftwo
 \else \expandafter \@secondoftwo
 \fi
}%
\providecommand \natexlab [1]{#1}%
\providecommand \enquote  [1]{``#1''}%
\providecommand \bibnamefont  [1]{#1}%
\providecommand \bibfnamefont [1]{#1}%
\providecommand \citenamefont [1]{#1}%
\providecommand \href@noop [0]{\@secondoftwo}%
\providecommand \href [0]{\begingroup \@sanitize@url \@href}%
\providecommand \@href[1]{\@@startlink{#1}\@@href}%
\providecommand \@@href[1]{\endgroup#1\@@endlink}%
\providecommand \@sanitize@url [0]{\catcode `\\12\catcode `\$12\catcode
  `\&12\catcode `\#12\catcode `\^12\catcode `\_12\catcode `\%12\relax}%
\providecommand \@@startlink[1]{}%
\providecommand \@@endlink[0]{}%
\providecommand \url  [0]{\begingroup\@sanitize@url \@url }%
\providecommand \@url [1]{\endgroup\@href {#1}{\urlprefix }}%
\providecommand \urlprefix  [0]{URL }%
\providecommand \Eprint [0]{\href }%
\providecommand \doibase [0]{https://doi.org/}%
\providecommand \selectlanguage [0]{\@gobble}%
\providecommand \bibinfo  [0]{\@secondoftwo}%
\providecommand \bibfield  [0]{\@secondoftwo}%
\providecommand \translation [1]{[#1]}%
\providecommand \BibitemOpen [0]{}%
\providecommand \bibitemStop [0]{}%
\providecommand \bibitemNoStop [0]{.\EOS\space}%
\providecommand \EOS [0]{\spacefactor3000\relax}%
\providecommand \BibitemShut  [1]{\csname bibitem#1\endcsname}%
\let\auto@bib@innerbib\@empty
\bibitem [{\citenamefont {Hollowood}(2013)}]{Hollowood2013}%
  \BibitemOpen
  \bibfield  {author} {\bibinfo {author} {\bibfnamefont {T.}~\bibnamefont
  {Hollowood}},\ }\href
  {http://link.springer.com/content/pdf/10.1007/978-3-642-36312-2.pdf} {\emph
  {\bibinfo {title} {{Renormalization Group and Fixed Points: In Quantum Field
  Theory}}}}\ (\bibinfo {year} {2013})\BibitemShut {NoStop}%
\bibitem [{\citenamefont {Kleinert}\ and\ \citenamefont
  {Schulte-Frohlinde}(2001)}]{Kleinert-Borel}%
  \BibitemOpen
  \bibfield  {author} {\bibinfo {author} {\bibfnamefont {H.}~\bibnamefont
  {Kleinert}}\ and\ \bibinfo {author} {\bibfnamefont {V.}~\bibnamefont
  {Schulte-Frohlinde}},\ }\href {https://doi.org/10.1142/4733} {\emph {\bibinfo
  {title} {{Critical Properties of {$\phi^4$} -Theories}}}}\ (\bibinfo
  {publisher} {WORLD SCIENTIFIC},\ \bibinfo {year} {2001})\BibitemShut
  {NoStop}%
\bibitem [{\citenamefont {Shalaby}(2020{\natexlab{a}})}]{abodispute}%
  \BibitemOpen
  \bibfield  {author} {\bibinfo {author} {\bibfnamefont {A.~M.}\ \bibnamefont
  {Shalaby}},\ }\bibfield  {title} {\bibinfo {title} {{{$\lambda$}-point
  anomaly in view of the seven-loop hypergeometric resummation for the critical
  exponent $\nu$ of the O(2) {$\phi^4$} model}},\ }\href
  {https://doi.org/10.1103/PhysRevD.102.105017} {\bibfield  {journal} {\bibinfo
   {journal} {Phys. Rev. D}\ }\textbf {\bibinfo {volume} {102}},\ \bibinfo
  {pages} {105017} (\bibinfo {year} {2020}{\natexlab{a}})}\BibitemShut
  {NoStop}%
\bibitem [{\citenamefont {Shalaby}(2021)}]{Abo-expon7}%
  \BibitemOpen
  \bibfield  {author} {\bibinfo {author} {\bibfnamefont {A.~M.}\ \bibnamefont
  {Shalaby}},\ }\bibfield  {title} {\bibinfo {title} {{Critical exponents of
  the {$O(N)$}-symmetric {$\phi ^4$} model from the {$\varepsilon ^7$}
  hypergeometric-Meijer resummation}},\ }\href
  {https://doi.org/10.1140/epjc/s10052-021-08884-5} {\bibfield  {journal}
  {\bibinfo  {journal} {Eur. Phys. J. C}\ }\textbf {\bibinfo {volume} {81}},\
  \bibinfo {pages} {87} (\bibinfo {year} {2021})},\ \Eprint
  {https://arxiv.org/abs/2005.12714} {arXiv:2005.12714} \BibitemShut {NoStop}%
\bibitem [{\citenamefont {Chester}\ \emph {et~al.}(2020)\citenamefont
  {Chester}, \citenamefont {Landry}, \citenamefont {Liu}, \citenamefont
  {Poland}, \citenamefont {Simmons-Duffin}, \citenamefont {Su},\ and\
  \citenamefont {Vichi}}]{dispute}%
  \BibitemOpen
  \bibfield  {author} {\bibinfo {author} {\bibfnamefont {S.~M.}\ \bibnamefont
  {Chester}}, \bibinfo {author} {\bibfnamefont {W.}~\bibnamefont {Landry}},
  \bibinfo {author} {\bibfnamefont {J.}~\bibnamefont {Liu}}, \bibinfo {author}
  {\bibfnamefont {D.}~\bibnamefont {Poland}}, \bibinfo {author} {\bibfnamefont
  {D.}~\bibnamefont {Simmons-Duffin}}, \bibinfo {author} {\bibfnamefont
  {N.}~\bibnamefont {Su}},\ and\ \bibinfo {author} {\bibfnamefont
  {A.}~\bibnamefont {Vichi}},\ }\bibfield  {title} {\bibinfo {title} {{Carving
  out OPE space and precise O(2) model critical exponents}},\ }\href
  {https://doi.org/10.1007/JHEP06(2020)142} {\bibfield  {journal} {\bibinfo
  {journal} {J. High Energy Phys.}\ }\textbf {\bibinfo {volume}
  {2020}}\bibfield  {number} {\bibinfo  {number} { (6)},\ \bibinfo {pages}
  {142}},\ }\Eprint {https://arxiv.org/abs/1912.03324} {arXiv:1912.03324}
  \BibitemShut {NoStop}%
\bibitem [{dis(2020)}]{dispute1}%
  \BibitemOpen
  \bibfield  {title} {\bibinfo {title} {{Conformal bootstrap and the
  $\lambda$-point specific heat experimental anomaly}},\ }\href
  {https://www.condmatjclub.org/?p=4037} {\bibfield  {journal} {\bibinfo
  {journal} {J. Club Condens. Matter Phys.}\ } (\bibinfo {year}
  {2020})}\BibitemShut {NoStop}%
\bibitem [{\citenamefont {Butera}\ and\ \citenamefont
  {Comi}(2002{\natexlab{a}})}]{Butera2002}%
  \BibitemOpen
  \bibfield  {author} {\bibinfo {author} {\bibfnamefont {P.}~\bibnamefont
  {Butera}}\ and\ \bibinfo {author} {\bibfnamefont {M.}~\bibnamefont {Comi}},\
  }\bibfield  {title} {\bibinfo {title} {{An on-line library of extended
  high-temperature expansions of basic observables for the spin-S Ising models
  on two- and three-dimensional lattices}},\ }\href
  {https://doi.org/10.1023/A:1019995830014} {\bibfield  {journal} {\bibinfo
  {journal} {J. Stat. Phys.}\ }\textbf {\bibinfo {volume} {109}},\ \bibinfo
  {pages} {311} (\bibinfo {year} {2002}{\natexlab{a}})}\BibitemShut {NoStop}%
\bibitem [{\citenamefont {Butera}\ and\ \citenamefont
  {Comi}(2002{\natexlab{b}})}]{Butera-PRB}%
  \BibitemOpen
  \bibfield  {author} {\bibinfo {author} {\bibfnamefont {P.}~\bibnamefont
  {Butera}}\ and\ \bibinfo {author} {\bibfnamefont {M.}~\bibnamefont {Comi}},\
  }\bibfield  {title} {\bibinfo {title} {{Critical universality and
  hyperscaling revisited for Ising models of general spin using extended
  high-temperature series}},\ }\href
  {https://doi.org/10.1103/PhysRevB.65.144431} {\bibfield  {journal} {\bibinfo
  {journal} {Phys. Rev. B}\ }\textbf {\bibinfo {volume} {65}},\ \bibinfo
  {pages} {144431} (\bibinfo {year} {2002}{\natexlab{b}})}\BibitemShut
  {NoStop}%
\bibitem [{\citenamefont {Hasenbusch}(2020)}]{MC20N3}%
  \BibitemOpen
  \bibfield  {author} {\bibinfo {author} {\bibfnamefont {M.}~\bibnamefont
  {Hasenbusch}},\ }\bibfield  {title} {\bibinfo {title} {{Monte Carlo study of
  a generalized icosahedral model on the simple cubic lattice}},\ }\href
  {https://doi.org/10.1103/PhysRevB.102.024406} {\bibfield  {journal} {\bibinfo
   {journal} {Phys. Rev. B}\ }\textbf {\bibinfo {volume} {102}},\ \bibinfo
  {pages} {024406} (\bibinfo {year} {2020})}\BibitemShut {NoStop}%
\bibitem [{\citenamefont {Xu}\ \emph {et~al.}(2019)\citenamefont {Xu},
  \citenamefont {Sun}, \citenamefont {Lv},\ and\ \citenamefont {Deng}}]{MC19L}%
  \BibitemOpen
  \bibfield  {author} {\bibinfo {author} {\bibfnamefont {W.}~\bibnamefont
  {Xu}}, \bibinfo {author} {\bibfnamefont {Y.}~\bibnamefont {Sun}}, \bibinfo
  {author} {\bibfnamefont {J.-P.}\ \bibnamefont {Lv}},\ and\ \bibinfo {author}
  {\bibfnamefont {Y.}~\bibnamefont {Deng}},\ }\bibfield  {title} {\bibinfo
  {title} {{High-precision Monte Carlo study of several models in the
  three-dimensional U(1) universality class}},\ }\href
  {https://doi.org/10.1103/PhysRevB.100.064525} {\bibfield  {journal} {\bibinfo
   {journal} {Phys. Rev. B}\ }\textbf {\bibinfo {volume} {100}},\ \bibinfo
  {pages} {064525} (\bibinfo {year} {2019})}\BibitemShut {NoStop}%
\bibitem [{\citenamefont {Hasenbusch}(2019)}]{MC19}%
  \BibitemOpen
  \bibfield  {author} {\bibinfo {author} {\bibfnamefont {M.}~\bibnamefont
  {Hasenbusch}},\ }\bibfield  {title} {\bibinfo {title} {{Monte Carlo study of
  an improved clock model in three dimensions}},\ }\href
  {https://doi.org/10.1103/PhysRevB.100.224517} {\bibfield  {journal} {\bibinfo
   {journal} {Phys. Rev. B}\ }\textbf {\bibinfo {volume} {100}},\ \bibinfo
  {pages} {224517} (\bibinfo {year} {2019})}\BibitemShut {NoStop}%
\bibitem [{\citenamefont {Kleinert}\ and\ \citenamefont
  {Schulte-Frohlinde}(1995)}]{Kleinert1995}%
  \BibitemOpen
  \bibfield  {author} {\bibinfo {author} {\bibfnamefont {H.}~\bibnamefont
  {Kleinert}}\ and\ \bibinfo {author} {\bibfnamefont {V.}~\bibnamefont
  {Schulte-Frohlinde}},\ }\bibfield  {title} {\bibinfo {title} {{Exact
  five-loop renormalization group functions of $\phi^4$-theory with
  $O(N)$-symmetric and cubic interactions. Critical exponents up to
  $\epsilon^5$}},\ }\href {https://doi.org/10.1016/0370-2693(94)01377-O}
  {\bibfield  {journal} {\bibinfo  {journal} {Phys. Lett. B}\ }\textbf
  {\bibinfo {volume} {342}},\ \bibinfo {pages} {284} (\bibinfo {year}
  {1995})}\BibitemShut {NoStop}%
\bibitem [{\citenamefont {Adzhemyan}\ \emph {et~al.}(2019)\citenamefont
  {Adzhemyan}, \citenamefont {Ivanova}, \citenamefont {Kompaniets},
  \citenamefont {Kudlis},\ and\ \citenamefont {Sokolov}}]{Adzhemyan2019}%
  \BibitemOpen
  \bibfield  {author} {\bibinfo {author} {\bibfnamefont {L.~T.}\ \bibnamefont
  {Adzhemyan}}, \bibinfo {author} {\bibfnamefont {E.~V.}\ \bibnamefont
  {Ivanova}}, \bibinfo {author} {\bibfnamefont {M.~V.}\ \bibnamefont
  {Kompaniets}}, \bibinfo {author} {\bibfnamefont {A.}~\bibnamefont {Kudlis}},\
  and\ \bibinfo {author} {\bibfnamefont {A.~I.}\ \bibnamefont {Sokolov}},\
  }\bibfield  {title} {\bibinfo {title} {{Six-loop $\epsilon$ expansion study
  of three-dimensional n-vector model with cubic anisotropy}},\ }\href
  {https://doi.org/10.1016/j.nuclphysb.2019.02.001} {\bibfield  {journal}
  {\bibinfo  {journal} {Nucl. Phys. B}\ }\textbf {\bibinfo {volume} {940}},\
  \bibinfo {pages} {332} (\bibinfo {year} {2019})},\ \Eprint
  {https://arxiv.org/abs/1901.02754} {arXiv:1901.02754} \BibitemShut {NoStop}%
\bibitem [{\citenamefont {Zinn-Justin}(2001)}]{zin-cr}%
  \BibitemOpen
  \bibfield  {author} {\bibinfo {author} {\bibfnamefont {J.}~\bibnamefont
  {Zinn-Justin}},\ }\bibfield  {title} {\bibinfo {title} {{Precise
  determination of critical exponents and equation of state by field theory
  methods}},\ }\href {https://doi.org/10.1016/S0370-1573(00)00126-5} {\bibfield
   {journal} {\bibinfo  {journal} {Phys. Rep.}\ }\textbf {\bibinfo {volume}
  {344}},\ \bibinfo {pages} {159} (\bibinfo {year} {2001})}\BibitemShut
  {NoStop}%
\bibitem [{\citenamefont {Kompaniets}\ and\ \citenamefont
  {Panzer}(2017)}]{Borel-6L}%
  \BibitemOpen
  \bibfield  {author} {\bibinfo {author} {\bibfnamefont {M.~V.}\ \bibnamefont
  {Kompaniets}}\ and\ \bibinfo {author} {\bibfnamefont {E.}~\bibnamefont
  {Panzer}},\ }\bibfield  {title} {\bibinfo {title} {{Minimally subtracted
  six-loop renormalization of $O(n)$-symmetric $\phi^4$ theory and critical
  exponents}},\ }\href {https://doi.org/10.1103/PhysRevD.96.036016} {\bibfield
  {journal} {\bibinfo  {journal} {Phys. Rev. D}\ }\textbf {\bibinfo {volume}
  {96}},\ \bibinfo {pages} {036016} (\bibinfo {year} {2017})}\BibitemShut
  {NoStop}%
\bibitem [{\citenamefont {Schnetz}(2018)}]{Schnetz2018}%
  \BibitemOpen
  \bibfield  {author} {\bibinfo {author} {\bibfnamefont {O.}~\bibnamefont
  {Schnetz}},\ }\bibfield  {title} {\bibinfo {title} {{Numbers and functions in
  quantum field theory}},\ }\href {https://doi.org/10.1103/PhysRevD.97.085018}
  {\bibfield  {journal} {\bibinfo  {journal} {Phys. Rev. D}\ }\textbf {\bibinfo
  {volume} {97}},\ \bibinfo {pages} {085018} (\bibinfo {year} {2018})},\
  \Eprint {https://arxiv.org/abs/1606.08598} {arXiv:1606.08598} \BibitemShut
  {NoStop}%
\bibitem [{\citenamefont {{Le Guillou}}\ and\ \citenamefont
  {Zinn-Justin}(1977)}]{LeGuillou1977}%
  \BibitemOpen
  \bibfield  {author} {\bibinfo {author} {\bibfnamefont {J.~C.}\ \bibnamefont
  {{Le Guillou}}}\ and\ \bibinfo {author} {\bibfnamefont {J.}~\bibnamefont
  {Zinn-Justin}},\ }\bibfield  {title} {\bibinfo {title} {{Critical exponents
  for the n-vector model in three dimensions from field theory}},\ }\href
  {https://doi.org/10.1103/PhysRevLett.39.95} {\bibfield  {journal} {\bibinfo
  {journal} {Phys. Rev. Lett.}\ }\textbf {\bibinfo {volume} {39}},\ \bibinfo
  {pages} {95} (\bibinfo {year} {1977})}\BibitemShut {NoStop}%
\bibitem [{\citenamefont {Shalaby}(2020{\natexlab{b}})}]{abo-precize}%
  \BibitemOpen
  \bibfield  {author} {\bibinfo {author} {\bibfnamefont {A.~M.}\ \bibnamefont
  {Shalaby}},\ }\bibfield  {title} {\bibinfo {title} {{Precise critical
  exponents of the $O(N)$-symmetric quantum field model using
  hypergeometric-Meijer resummation}},\ }\href
  {https://doi.org/10.1103/PhysRevD.101.105006} {\bibfield  {journal} {\bibinfo
   {journal} {Phys. Rev. D}\ }\textbf {\bibinfo {volume} {101}},\ \bibinfo
  {pages} {105006} (\bibinfo {year} {2020}{\natexlab{b}})}\BibitemShut
  {NoStop}%
\bibitem [{\citenamefont {Onsager}(1944)}]{Onsager}%
  \BibitemOpen
  \bibfield  {author} {\bibinfo {author} {\bibfnamefont {L.}~\bibnamefont
  {Onsager}},\ }\bibfield  {title} {\bibinfo {title} {{Crystal Statistics. I. A
  Two-Dimensional Model with an Order-Disorder Transition}},\ }\href
  {https://doi.org/10.1103/PhysRev.65.117} {\bibfield  {journal} {\bibinfo
  {journal} {Phys. Rev.}\ }\textbf {\bibinfo {volume} {65}},\ \bibinfo {pages}
  {117} (\bibinfo {year} {1944})}\BibitemShut {NoStop}%
\bibitem [{\citenamefont {Nienhuis}(1982)}]{Nienhuis}%
  \BibitemOpen
  \bibfield  {author} {\bibinfo {author} {\bibfnamefont {B.}~\bibnamefont
  {Nienhuis}},\ }\bibfield  {title} {\bibinfo {title} {{Exact Critical Point
  and Critical Exponents of $O(n)$ Models in Two Dimensions}},\ }\href
  {https://doi.org/10.1103/PhysRevLett.49.1062} {\bibfield  {journal} {\bibinfo
   {journal} {Phys. Rev. Lett.}\ }\textbf {\bibinfo {volume} {49}},\ \bibinfo
  {pages} {1062} (\bibinfo {year} {1982})}\BibitemShut {NoStop}%
\bibitem [{\citenamefont {Lipa}\ \emph {et~al.}(2003)\citenamefont {Lipa},
  \citenamefont {Nissen}, \citenamefont {Stricker}, \citenamefont {Swanson},\
  and\ \citenamefont {Chui}}]{experiment}%
  \BibitemOpen
  \bibfield  {author} {\bibinfo {author} {\bibfnamefont {J.~A.}\ \bibnamefont
  {Lipa}}, \bibinfo {author} {\bibfnamefont {J.~A.}\ \bibnamefont {Nissen}},
  \bibinfo {author} {\bibfnamefont {D.~A.}\ \bibnamefont {Stricker}}, \bibinfo
  {author} {\bibfnamefont {D.~R.}\ \bibnamefont {Swanson}},\ and\ \bibinfo
  {author} {\bibfnamefont {T.~C.~P.}\ \bibnamefont {Chui}},\ }\bibfield
  {title} {\bibinfo {title} {{Specific heat of liquid helium in zero gravity
  very near the lambda point}},\ }\href
  {https://doi.org/10.1103/PhysRevB.68.174518} {\bibfield  {journal} {\bibinfo
  {journal} {Phys. Rev. B}\ }\textbf {\bibinfo {volume} {68}},\ \bibinfo
  {pages} {174518} (\bibinfo {year} {2003})}\BibitemShut {NoStop}%
\bibitem [{\citenamefont {Kleinert}(1999)}]{Kleinert-strong1}%
  \BibitemOpen
  \bibfield  {author} {\bibinfo {author} {\bibfnamefont {H.}~\bibnamefont
  {Kleinert}},\ }\bibfield  {title} {\bibinfo {title} {{Critical exponents from
  seven-loop strong-coupling $\phi^4$theory in three dimensions}},\ }\href
  {https://doi.org/10.1103/PhysRevD.60.085001} {\bibfield  {journal} {\bibinfo
  {journal} {Phys. Rev. D}\ }\textbf {\bibinfo {volume} {60}},\ \bibinfo
  {pages} {085001} (\bibinfo {year} {1999})}\BibitemShut {NoStop}%
\bibitem [{\citenamefont {Kleinert}(1998)}]{keinert-strong}%
  \BibitemOpen
  \bibfield  {author} {\bibinfo {author} {\bibfnamefont {H.}~\bibnamefont
  {Kleinert}},\ }\bibfield  {title} {\bibinfo {title} {{Strong-coupling
  behavior of {$\phi^4$} theories and critical exponents}},\ }\href
  {https://doi.org/10.1103/PhysRevD.57.2264} {\bibfield  {journal} {\bibinfo
  {journal} {Phys. Rev. D}\ }\textbf {\bibinfo {volume} {57}},\ \bibinfo
  {pages} {2264} (\bibinfo {year} {1998})}\BibitemShut {NoStop}%
\bibitem [{\citenamefont {Kleinert}(1995)}]{variational95}%
  \BibitemOpen
  \bibfield  {author} {\bibinfo {author} {\bibfnamefont {H.}~\bibnamefont
  {Kleinert}},\ }\bibfield  {title} {\bibinfo {title} {{Variational
  interpolation algorithm between weak- and strong-coupling expansions —
  application to the polaron}},\ }\href
  {https://doi.org/10.1016/0375-9601(95)00683-T} {\bibfield  {journal}
  {\bibinfo  {journal} {Phys. Lett. A}\ }\textbf {\bibinfo {volume} {207}},\
  \bibinfo {pages} {133} (\bibinfo {year} {1995})}\BibitemShut {NoStop}%
\bibitem [{\citenamefont {Janke}\ and\ \citenamefont
  {Kleinert}(1995)}]{Janke95}%
  \BibitemOpen
  \bibfield  {author} {\bibinfo {author} {\bibfnamefont {W.}~\bibnamefont
  {Janke}}\ and\ \bibinfo {author} {\bibfnamefont {H.}~\bibnamefont
  {Kleinert}},\ }\bibfield  {title} {\bibinfo {title} {{Convergent
  Strong-Coupling Expansions from Divergent Weak-Coupling Perturbation
  Theory}},\ }\href {https://doi.org/10.1103/PhysRevLett.75.2787} {\bibfield
  {journal} {\bibinfo  {journal} {Phys. Rev. Lett.}\ }\textbf {\bibinfo
  {volume} {75}},\ \bibinfo {pages} {2787} (\bibinfo {year}
  {1995})}\BibitemShut {NoStop}%
\bibitem [{\citenamefont {Peskin}\ and\ \citenamefont
  {Schroeder}(1995)}]{Peskin}%
  \BibitemOpen
  \bibfield  {author} {\bibinfo {author} {\bibfnamefont {M.}~\bibnamefont
  {Peskin}}\ and\ \bibinfo {author} {\bibfnamefont {D.}~\bibnamefont
  {Schroeder}},\ }\href@noop {} {\emph {\bibinfo {title} {{An Introduction to
  quantum field theory}}}}\ (\bibinfo  {publisher} {Addison-Wesley},\ \bibinfo
  {address} {Reading, USA},\ \bibinfo {year} {1995})\BibitemShut {NoStop}%
\bibitem [{\citenamefont {Stevenson}(1984)}]{GEPQM}%
  \BibitemOpen
  \bibfield  {author} {\bibinfo {author} {\bibfnamefont {P.~M.}\ \bibnamefont
  {Stevenson}},\ }\bibfield  {title} {\bibinfo {title} {{Gaussian effective
  potential: Quantum mechanics}},\ }\href
  {https://doi.org/10.1103/PhysRevD.30.1712} {\bibfield  {journal} {\bibinfo
  {journal} {Phys. Rev. D}\ }\textbf {\bibinfo {volume} {30}},\ \bibinfo
  {pages} {1712} (\bibinfo {year} {1984})}\BibitemShut {NoStop}%
\bibitem [{\citenamefont {Jasch}\ and\ \citenamefont
  {Kleinert}(2001)}]{harm-sigma}%
  \BibitemOpen
  \bibfield  {author} {\bibinfo {author} {\bibfnamefont {F.}~\bibnamefont
  {Jasch}}\ and\ \bibinfo {author} {\bibfnamefont {H.}~\bibnamefont
  {Kleinert}},\ }\bibfield  {title} {\bibinfo {title} {{Fast-convergent
  resummation algorithm and critical exponents of $\phi^4$-theory in three
  dimensions}},\ }\href {https://doi.org/10.1063/1.1289377} {\bibfield
  {journal} {\bibinfo  {journal} {J. Math. Phys.}\ }\textbf {\bibinfo {volume}
  {42}},\ \bibinfo {pages} {52} (\bibinfo {year} {2001})}\BibitemShut {NoStop}%
\bibitem [{\citenamefont {Sk{\'{a}}la}\ \emph {et~al.}(1999)\citenamefont
  {Sk{\'{a}}la}, \citenamefont {C{\'{i}}zek},\ and\ \citenamefont
  {Zamastil}}]{Large-strong1}%
  \BibitemOpen
  \bibfield  {author} {\bibinfo {author} {\bibfnamefont {L.}~\bibnamefont
  {Sk{\'{a}}la}}, \bibinfo {author} {\bibfnamefont {J.}~\bibnamefont
  {C{\'{i}}zek}},\ and\ \bibinfo {author} {\bibfnamefont {J.}~\bibnamefont
  {Zamastil}},\ }\bibfield  {title} {\bibinfo {title} {{Strong coupling
  perturbation expansions for anharmonic oscillators. Numerical results}},\
  }\href {https://doi.org/10.1088/0305-4470/32/30/314} {\bibfield  {journal}
  {\bibinfo  {journal} {J. Phys. A. Math. Gen.}\ }\textbf {\bibinfo {volume}
  {32}},\ \bibinfo {pages} {5715} (\bibinfo {year} {1999})}\BibitemShut
  {NoStop}%
\bibitem [{\citenamefont {Shalaby}(2009)}]{x4pabo}%
  \BibitemOpen
  \bibfield  {author} {\bibinfo {author} {\bibfnamefont {A.~M.}\ \bibnamefont
  {Shalaby}},\ }\bibfield  {title} {\bibinfo {title} {{Effective field
  calculations of the energy spectrum of the $\mathcal{PT}$-symmetrix $-x^4$
  potential}},\ }\href {https://doi.org/10.1103/PhysRevD.79.065017} {\bibfield
  {journal} {\bibinfo  {journal} {Phys. Rev. D}\ }\textbf {\bibinfo {volume}
  {79}},\ \bibinfo {pages} {065017} (\bibinfo {year} {2009})}\BibitemShut
  {NoStop}%
\bibitem [{\citenamefont {Jones}\ and\ \citenamefont {Mateo}(2006)}]{Jonesx4}%
  \BibitemOpen
  \bibfield  {author} {\bibinfo {author} {\bibfnamefont {H.~F.}\ \bibnamefont
  {Jones}}\ and\ \bibinfo {author} {\bibfnamefont {J.}~\bibnamefont {Mateo}},\
  }\bibfield  {title} {\bibinfo {title} {{Equivalent Hermitian Hamiltonian for
  the non-Hermitian $-x^4$ potential}},\ }\href
  {https://doi.org/10.1103/PhysRevD.73.085002} {\bibfield  {journal} {\bibinfo
  {journal} {Phys. Rev. D}\ }\textbf {\bibinfo {volume} {73}},\ \bibinfo
  {pages} {085002} (\bibinfo {year} {2006})}\BibitemShut {NoStop}%
\bibitem [{\citenamefont {Coleman}(1975)}]{normalc}%
  \BibitemOpen
  \bibfield  {author} {\bibinfo {author} {\bibfnamefont {S.}~\bibnamefont
  {Coleman}},\ }\bibfield  {title} {\bibinfo {title} {{Quantum sine-Gordon
  equation as the massive Thirring model}},\ }\href
  {https://doi.org/10.1103/PhysRevD.11.2088} {\bibfield  {journal} {\bibinfo
  {journal} {Phys. Rev. D}\ }\textbf {\bibinfo {volume} {11}},\ \bibinfo
  {pages} {2088} (\bibinfo {year} {1975})}\BibitemShut {NoStop}%
\bibitem [{\citenamefont {Shalaby}(2007)}]{Shalaby2007}%
  \BibitemOpen
  \bibfield  {author} {\bibinfo {author} {\bibfnamefont {A.}~\bibnamefont
  {Shalaby}},\ }\bibfield  {title} {\bibinfo {title} {{Non-perturbative
  calculations for the effective potential of the $\mathcal{PT}$- symmetric and
  non-Hermitian $(-g\phi^4)$ field theoretical model}},\ }\href
  {https://doi.org/10.1140/epjc/s10052-007-0236-4} {\bibfield  {journal}
  {\bibinfo  {journal} {Eur. Phys. J. C}\ }\textbf {\bibinfo {volume} {50}},\
  \bibinfo {pages} {999} (\bibinfo {year} {2007})}\BibitemShut {NoStop}%
\bibitem [{\citenamefont {Din}(1971)}]{normal2}%
  \BibitemOpen
  \bibfield  {author} {\bibinfo {author} {\bibfnamefont {A.~M.}\ \bibnamefont
  {Din}},\ }\bibfield  {title} {\bibinfo {title} {{Some Remarks on the Normal
  Ordering of Lagrangians}},\ }\href {https://doi.org/10.1103/PhysRevD.4.995}
  {\bibfield  {journal} {\bibinfo  {journal} {Phys. Rev. D}\ }\textbf {\bibinfo
  {volume} {4}},\ \bibinfo {pages} {995} (\bibinfo {year} {1971})}\BibitemShut
  {NoStop}%
\bibitem [{\citenamefont {Stevenson}(1985)}]{GEPII}%
  \BibitemOpen
  \bibfield  {author} {\bibinfo {author} {\bibfnamefont {P.~M.}\ \bibnamefont
  {Stevenson}},\ }\bibfield  {title} {\bibinfo {title} {{Gaussian effective
  potential. II. $\lambda \phi^4$ field theory}},\ }\href
  {https://doi.org/10.1103/PhysRevD.32.1389} {\bibfield  {journal} {\bibinfo
  {journal} {Phys. Rev. D}\ }\textbf {\bibinfo {volume} {32}},\ \bibinfo
  {pages} {1389} (\bibinfo {year} {1985})}\BibitemShut {NoStop}%
\bibitem [{\citenamefont {Shalaby}(2022)}]{universal}%
  \BibitemOpen
  \bibfield  {author} {\bibinfo {author} {\bibfnamefont {A.~M.}\ \bibnamefont
  {Shalaby}},\ }\bibfield  {title} {\bibinfo {title} {{Universal large-order
  asymptotic behavior of the strong-coupling and high-temperature series
  expansions}},\ }\href {https://doi.org/10.1103/PhysRevD.105.045004}
  {\bibfield  {journal} {\bibinfo  {journal} {Phys. Rev. D}\ }\textbf {\bibinfo
  {volume} {105}},\ \bibinfo {pages} {045004} (\bibinfo {year} {2022})},\
  \Eprint {https://arxiv.org/abs/1911.03571} {arXiv:1911.03571} \BibitemShut
  {NoStop}%
\end{thebibliography}%

\end{document}